\documentclass{elsart}
\usepackage{epsfig}
\newcommand{\lit}{\scriptscriptstyle}

\begin{document}

\begin{frontmatter}

\title{Modeling of the saturation current of a fission chamber taking into account the distorsion of electric field due to space charge effects}

\author[ens]{O. Poujade}
\author[cea]{A. Lebrun\thanksref{cor}}

\address[ens]{Ecole Normale Sup\'erieure de Cachan, D\'epartement de physique, F-94235 Cachan, France}

\address[cea]{CEA Cadarache, DRN/DER/SSAE/LSMR, F-13108 St Paul-lez-Durance, France}

\thanks[cor]{corresponding author. e-mail : alain.lebrun@cea.fr}

\begin{abstract}

Fission chambers were first made fifty years ago for neutron detection. At the moment, the French Atomic Energy Commission  \textsf{(CEA-Cadarache)} is developing a sub-miniature fission chamber technology with a diameter of 1.5 mm working in the current mode \cite{Bign}. To be able to measure intense fluxes, it is necessary to adjust the chamber geometry and the gas pressure before testing it under real neutron flux. In the present paper, we describe a theoretical method to foresee the current-voltage characteristics (sensitivity and saturation plateau) of a fission chamber whose geometrical features are given, taking into account the neutron flux to be measured (spectrum and intensity). The proposed theoretical model describes electric field distortion resulting from charge collection effect. A computer code has been developed on this model basis. Its application to 3 kinds of fission chambers indicates excellent agreement between theoretical model and measured characteristics. 

\end{abstract}


\end{frontmatter}

\section{INTRODUCTION}

A fission chamber is a device that measures a neutron flux and that is weakly disturbed by the gamma flux \cite[p 503]{Knoll}.
\\
In fig. 1, we show the basic principle behind the use of a fission chamber as a neutron detector. A thin layer of fissile material (\nuc{235}{U} for instance) is deposited on the anode. When a neutron traverses the detector, it may collide with a fissile atom (the probability is proportional to the fission reaction cross section) and the resulting reaction generates two charged particles (Fission Fragment :\textsf{ FF}) produced by fission. Those charged particles cross the argon atmosphere between the two electrodes of the chamber (anode and cathode) and ionize the gas on their trajectory creating an average of $\mathsf{0.3~ion{\lit{-}}pairs/nm}$ ($\mathsf{Ar}^{\lit{+}}$, $\mathsf{e}^{\lit{-}}$). A \textsf{DC} voltage of a few hundred volts being applied between the two electrodes, those electrons and ions drift across the gas in response to the applied electric field producing a signal that can be amplified and processed. 
\\
The voltage-current characteristic of a fission chamber can present a saturation plateau. It means that, in a  voltage interval, the current delivered by the chamber is roughly constant. The saturation current is proportional to the neutron flux, that is the reason why we need to know when the chamber works in saturation conditions. But, for a given geometry, the so called voltage interval depends on the neutron flux. The main aim of this study is to determine the relevant parameters of the chamber and the optimal voltage that answer the neutron flux requirement of the user. 
\\
Firstly, we will study quantitatively the conversion of neutrons into fission fragments and into ion-pairs. Secondly, we will spend some time to understand the charge collection and the creation of the signal. Then, we will compare the results predicted by our model with those obtained experimentally.  

\begin{figure}
\epsfxsize=7cm
\epsfysize=6cm
\centerline{\epsfbox{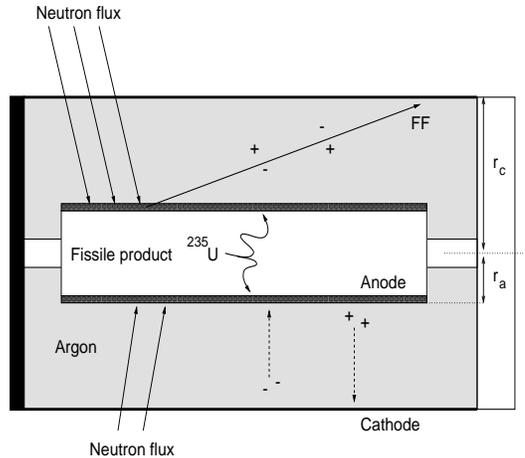}}
\caption{A schematic view of a fission chamber.}
\label{scefiss}
\end{figure}

\section{ION-PAIR CREATION}

\subsection{Neutrons make Fissions}

We are dealing with the fission reaction that turns an\- un\-char\-ged particle (neutron) into charged par\-ti\-cles (\textsf{FF}) which are easier to detect~: $$(\mathsf{n},\mathsf{U}^{235})\longrightarrow(\mathsf{2FF,2.3 neutrons,energy})$$ Let $\mu_{s}$ be the surface mass of fissile product (for a subminiature fission chamber $\mathsf{\mu_{s}=760\,\mu g\cdot cm^{-2}}$). This parameter decreases during the life of the chamber, because fissile atoms are not regenerated. This is our stock of fission fragments. The rate of fission reactions per surface unit is given by $$\mathsf{ N_{fst}=\int_{0}^{\infty}\sigma(\epsilon)\,\frac{\mu_{s}}{ M_{\lit{U}}} N_{\lit{A}}\Phi_{\epsilon}(\epsilon)\,d\epsilon}$$ (the subscript \underline{\textsf{fst}} stands for {\underline{\bf f}}ission per unit {\underline{\bf s}}urface per unit {\underline{\bf t}}ime). $\mathsf{ M_{U}}$ is the molar mass of \nuc{235}{U}, $\mathsf{ N_{\lit{A}}}$ is the number of Avogadro, $\mathsf{\sigma(\epsilon)}$ is the fission cross section for neutron of energy $\epsilon$ and $\mathsf{\Phi_{\epsilon}(\epsilon)}$ is the energetic density of the incoming neutron flux. For the next steps we will call $\mathsf{\chi=\int_{0}^{\infty}\sigma(\epsilon)\Phi_{\epsilon}(\epsilon)\,d\epsilon}$. A same value of $\mathsf{\chi}$ can be generated by various neutron spectra. We just have to find the right spectra of incoming neutron $\mathsf{\Phi_{\epsilon}(\epsilon)}$. This could cause a problem as we are going to see in the next part.

\subsection{Fissions make fission fragments}

For example : take two different spectra of incoming neutrons that generate the same fission rate $\mathsf{\chi}$. Let us consider one flux of incoming neutrons of energy $\mathsf{0.04~eV}$ and one of ener\-gy $\mathsf{1~MeV}$. For \nuc{235}{U}, $$\mathsf{\sigma(0.04~eV)=570~barn}~and~\mathsf{\sigma(1~MeV)=0.1~barn.}$$ Then a flux of $\mathsf{1.0~10^{12}n\cdot cm^{-2}\cdot s^{-1}}$ for $\mathsf{0.04~eV}$ neutrons and a flux of about $\mathsf{5.7~10^{15}n\cdot cm^{-2}\cdot s^{-1}}$ for $\mathsf{1~MeV}$ neutrons result in the same reaction rate $\chi $  : $\mathsf{\chi=5.7~10^{-10}\,s^{-1}}$ and the same number of fission fragments is created. The problem is that, a priori, the nature and the energy of these fragments is not the same as the ionization rate in the gas. Fortunately, the distribution of fission fragments (fig. 2) do not vary significantly with the energy of the incoming neutrons. Moreover, since the average energy of the 2 $\mathsf{FF}$ is about $\mathsf{160~MeV}$, we can say that even a difference of $\mathsf{15~MeV}$ in the energy of incoming neutron induces an error of less than $\mathsf{10~\%}$. So, we will suppose that the fission fragments distribution and the fission fragments energy distribution do not vary significantly for incoming neutron energies between $\mathsf{0~MeV}$ and $\mathsf{15~MeV}$.

\begin{figure}[!th]
\epsfxsize=7cm
\epsfysize=7cm
\centerline{\epsfbox{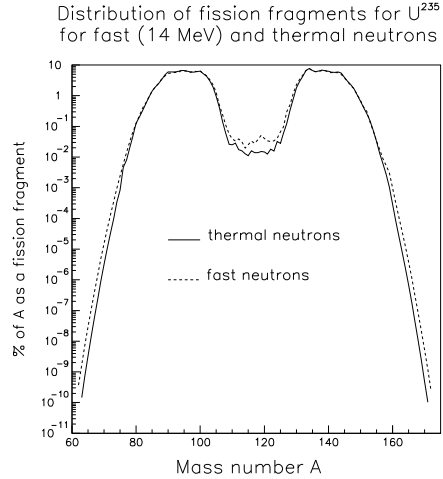}}
\caption{distribution of \nuc{235}{U} fission fragments.}
\label{dist}
\end{figure}

\subsection{Fission fragments make ion-pairs}

One fission gives birth to two fission fragments. Their average energy is around $\mathsf{90~MeV}$ for the lightest and $\mathsf{60~MeV}$ for the heaviest. If the energy of the incident neutron is negligible compared to $\mathsf{90~MeV}$ and $\mathsf{60~MeV}$ it is accurate to say those FF emerge in opposite directions. Thus only one FF participates in the gas ionization, the other one being absorbed by the anode.

\subsubsection{Calculation of the number of ion-pairs created per unit time per unit volume}

We want to calculate the density of ion-pairs created per unit time ($\mathsf{ N(r)}$) at a distance $\mathsf{r}$ from the anode axis (see on fig. 3). We assume the hypothesis that FFs are emitted in every direction with the same probability. Let $\mathsf{ N_{ft}}$ be the number of fission reactions on the surface element $\mathsf{dS}$ per unit time. $$ \mathsf{ N_{ft}= N_{fst}\,dz\,r_{a}\,d\theta}$$ Let $\mathsf{I^{c}(x)}$ be the average number of ion-pairs created by a fission fragment per unit of length travelled per unit pressure. $\mathsf{x}$ representes the distance travelled from the creation point. We can calculate the number of ion-pairs created per unit of time in the volume element $\mathsf{d\tau=r^{\prime}d\psi\,dr^{\prime}dz}$ around the point $\mathsf{P}$. This number is $$\mathsf{dN= N_{ft}\,cos(\psi)\frac{d\psi}{\pi}pI^{c}(r^{\prime})dr^{\prime}}$$ where \textsf{p} is the pressure of the chamber. Thus, the density of ion-pairs created in $\mathsf{P}$ because of $\mathsf{dS}$ is $$\mathsf{\frac{dN}{d\tau}=\frac{ N_{fst}}{\pi}\frac{r_{a}}{r^{\prime}}cos(\psi)d\theta\,pI^{c}(r^{\prime})}$$ In the case of a sub-miniature fission chamber the trajectory of fission fragments is short enough to consider $\mathsf{I^{c}(r)}$ to be constant equal to $\mathsf{I^{c}_{0}}$ Thus we can say that the density of ion-pairs created at distance $\mathsf{r}$ from the axis per unit time is 
\begin{equation}
\mathsf{N(r)=\int_{-acos(r_{a}/r)}^{acos(r_{a}/r)}{N_{fst}}\frac{pI^{c}_{0}r_{a}(r\,cos(\theta)-r_{a})}{r^{2}+r_{a}^{2}-2r\,r_{a}\,cos(\theta)}d\theta}
\end{equation}

\begin{figure}[!th]
\epsfxsize=5cm
\epsfysize=7cm
\centerline{\epsfbox{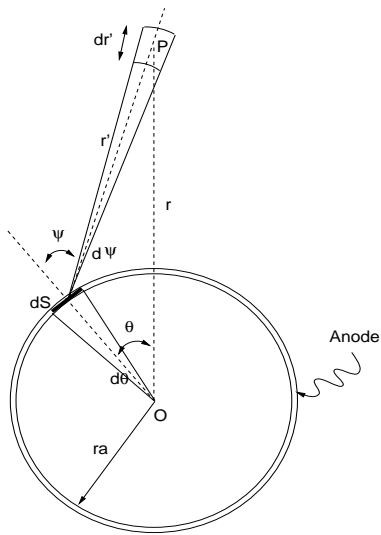}}
\caption{Schematic cross section of the Anode.}
\label{prof}
\end{figure}

\subsubsection{Justification of $\mathsf{I^{c}(r)}$ constant equal to $\mathsf{I^{c}_{0}}$}

The results of this part are due to the program SRIM of J.F Ziegler which calculates the stopping and range of ions in solids, liquids and gases \cite{Zieg}. In fig. 4, we show the energy loss of two typical fission fragments : \nuc{95}{Mo} at $\mathsf{90~MeV}$ and \nuc{139}{La} at $\mathsf{60~MeV}$. The space separating the electrodes of all types of fission chambers is no longer than a few milimeters. For such a distance, we can notice that the energy loss by ionization in argon is roughly constant equal to $\mathsf{4.5~keV.\mu m^{-1}}$ and so is $\mathsf{I^{c}(r)}$. Suppose that the creation of one ion-pair requires $\mathsf{23~eV}$ (ionization energy of argon). We can say that the number of ion-pairs created by one typical FF per unit of length through the gas is $\mathsf{I^{c}_{0}=2.\,10^{5}~ion{\lit{-}}pairs/mm/bar}$. This hypothesis will be the first source of error.

\begin{figure}[!ht]
\epsfxsize=7cm
\epsfysize=7cm
\centerline{\epsfbox{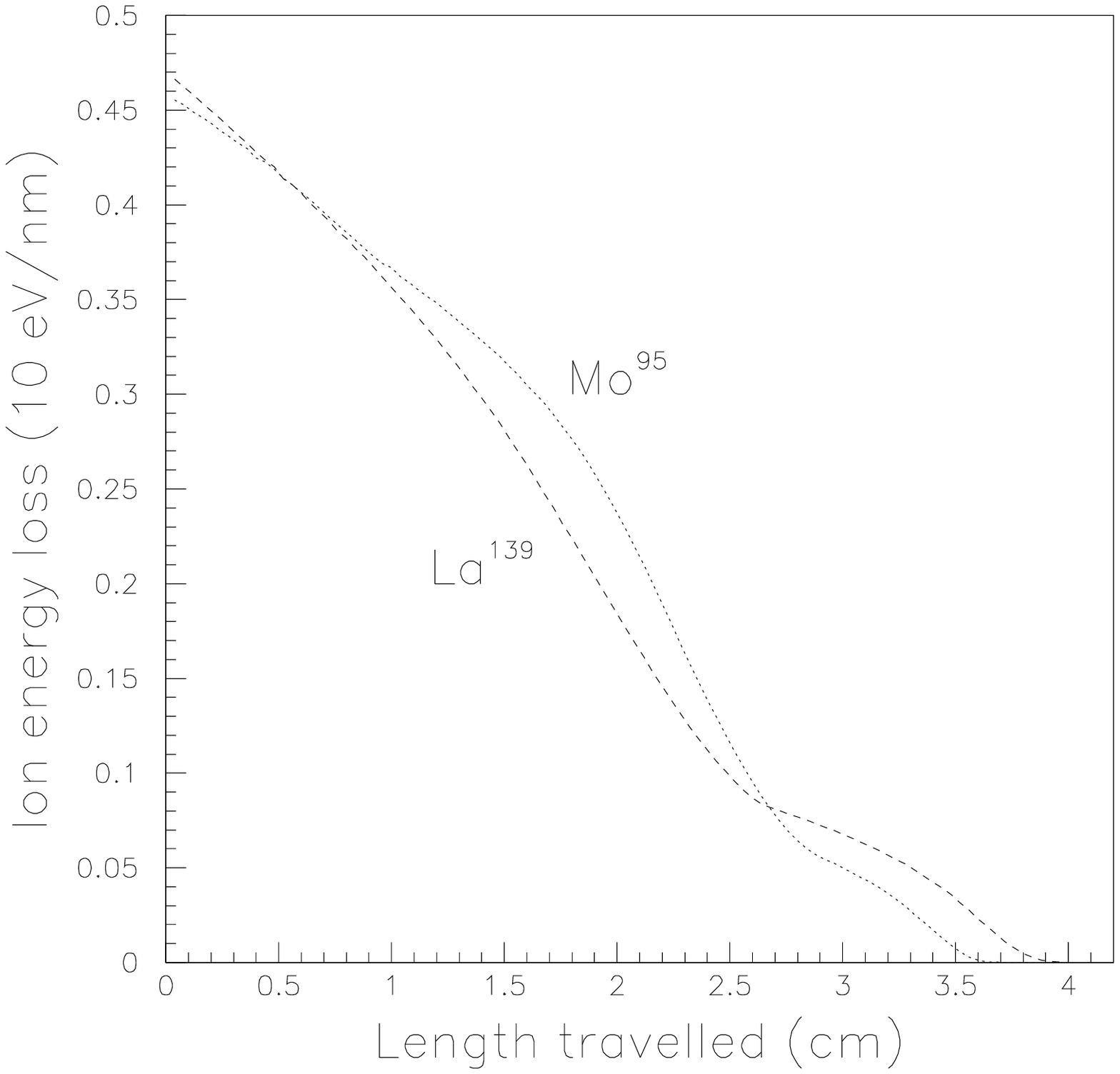}}
\caption{Energy loss for \nuc{95}{Mo} and \nuc{139}{La} in $\mathsf{1.0~bar}$ Argon.}
\label{ioniz}
\end{figure}

\subsubsection{Asymptotic hypothesis}

It would be quite difficult to work with (1). It simplifies singularily if, instead, we take the asymptotic version $\mathsf{(r\longrightarrow\,r_{a})}$ or $\mathsf{(r\longrightarrow\,+\infty)}$. Suppose $\mathsf{(r_{c}-r_{a})\ll r_{a}}$ then $\mathsf{r}$ remains in the vicinity of $\mathsf{r_{a}}$ and we can assume that $\mathsf{N(r)}$ is constant and equal to its value in $\mathsf{r_{a}}$
$$\mathsf{ N(r)=N_{0}=\frac{\pi}{2}N_{fst}pI^{c}_{0}}$$  Suppose, now, $\mathsf{(r_{c}-r_{a})\gg r_{a}}$ then we can take with a good accuracy the equivalent of $\mathsf{N(r)}$ at infinite, $$\mathsf{N(r)=N_{fst}pI^{c}_{0}\frac{r_{a}}{r}=\frac{\beta}{r}}$$  In fact, the first approximation fits with the case of plane electrodes, and the second squares with cylindrical geometry of the chamber. Those hypothesis will be the second source of error. 

\section{PHYSICAL PHENOMENA IN THE GAS}

Let $\mathsf{ N(r)}$ be the number of ion-pairs created by fission fragments per unit of time and volume in the gas (it depends on $\mathsf{\chi=\int\sigma(\epsilon)\Phi_{\epsilon}(\epsilon)d\epsilon}$ as seen before). We will suppose this number to be equal to $\mathsf{N_{0}}$ in the entire volume of the gas, because most fission chambers verify the geometrical hypothesis $\mathsf{(r_{c}-r_{a})\ll r_{a}}$. This approximation will give us only an order of magnitude. But, from an experimental point of view, it would be utopic to expect better than this. Indeed, the geometry of a fission chamber varies from one to the other. It is impossible to get a perfectly cylindrical geometry, and this may cause some difference of charge collection between two similar models of fission chambers. Various physical phenomena can occur in the gas \cite{Davi}. Comparing experiment and theory, we noticed that only certain effects could be taken into account. 

\subsection{Primary ionization}

The {\it{primary ionization}} corresponds to the ionization reaction\linebreak $\mathsf{(FF,Ar)\longrightarrow(FF,Ar^{\lit{+}},e^{\lit{-}})}$. The fission fragments ionize the argon and generate ion pairs ($\mathsf{Ar^{\lit{+}}}$,$\mathsf{e^{\lit{-}}}$) and they lose energy by this process (see on fig. 4).

\subsection{Electrical drift and temperature}

The DC voltage applied between the two electrodes generates an electric field that is responsible for the movement of electrons towards the anode and ions towards the cathode. This is the {\it{electrical drift}}. The respective velocity of electrons and ions in such a field are $\mathsf{\mu_{\lit{-}}\, E}$ and $\mathsf{\mu_{\lit{+}}\, E}$ where $\mu_{\lit{-}}$ and $\mu_{\lit{+}}$ are the mobility of those particles in the gas . In a gas, the mobility of a charged particle is assumed to be inversely proportional to the pressure of the gas \cite{Davi}. For argon : $$\mathsf{\mu_{\lit{-}}=0.13\,m^{2}.bar.V^{-1}.s^{-1}}$$
 $$\mathsf{\mu_{\lit{+}}=0.13\,10^{-3}m^{2}.bar.V^{-1}.s^{-1}}$$ Since the pressure is proportional to the temperature, we can say that the mobility varies as the inverse of the temperature. This will be the only effect of the temperature modelized. 

\subsection{Recombination}

Since $\mathsf{Ar^{\lit{+}}}$ and $\mathsf{e^{\lit{-}}}$ are oppositely charged particles, it is also possible that a certain amount of those particles recombine into mother particles $\mathsf{(e^{\lit{-}},Ar^{\lit{+}})\longrightarrow\, Ar}$. This is the {\it{recombination}} reaction \cite{Sail}. The rate of recombination depends on the local density of ions and electrons and is given by $$\mathsf{(\frac{dn_{\lit{-}}}{dt})_{rec}=(\frac{dn_{\lit{+}}}{dt})_{rec}=\alpha\, n_{\lit{-}}\, n_{\lit{+}}}$$ The tabulated value of $\mathsf{\alpha}$ is assumed to be $\mathsf{10^{-12}~SI}$ for argon

\subsection{Threshold condition}

Now, suppose that the electric field is strong enough to give the free electrons a kinetic energy of a few \textsf{eV} (order of magnitude). The {\it{secondary ionization}} reaction $\mathsf{(e^{\lit{-}},Ar)\longrightarrow(2e^{\lit{-}},Ar^{\lit{+}})}$ is then possible. The thermal kinetic energy is about $\mathsf{0.04~eV}$. It means that the kinetic energy of a few \textsf{eV} of electrons is mainly due to the electric field : $\mathsf{Kinetic~energy=\frac{1}{2}m_{e}(\mu_{\lit{-}}E_{S}/p)^{2}}$ is about a few \textsf{eV}. For a $\mathsf{1.0~bar}$ pressure of Argon we can calculate the electric field threshold ($\mathsf{E_{S}}$) that per\-mits the {\it{secondary ioniza\-tion}} reaction : $$\mathsf{E_{S}=some~units~of~10^{6}\,V\cdot m^{-1}}$$ In this way, this hypothesis can explain the fact that the end voltage of the saturation plateau is higher when the pressure is higher (ref. part $\mathsf{2.1}$). Indeed, in order to maintain the kinetic energy constant, we must keep the ratio $\mathsf{E_{S}/p}$ constant. It means that $\mathsf{E_{S}}$ is proportional to the pressure of argon. In our model, we took, for a $\mathsf{1.0~bar}$ chamber, $\mathsf{E_{S}=1.7\,10^{6}\,V\cdot m^{-1}}$.

\section{CHARGE COLLECTION}

\subsection{Minimum voltage for steady state solution}

The relevant signal of a fission chamber in the current mode is the \textsf{RMS} value of the current. We are not interested in the noise generated by the chamber. Thus, we are going to study the stationary state of the chamber. If we want a steady state solution to exist, all phenomena must balance. Suppose that the only effects we take into account are the primary ionization ( this is the dominant effect on the saturation plateau ) and the electrical drift. As soon as {\it recombination} is of the order of magnitude of primary ionization this hypothesis will not be valid anymore. These hypotheses will enable us to  compute easily the electron density, the ion density and the electric field. We expect a minimum value of the voltage under which the electric field is not strong enough to evacuate the charges created by the primary ionization. Thus, below this minimum voltage a steady state solution could not exist. We will evaluate the validity domain of these hypotheses by comparing the value of $\mathsf{\alpha n_{\lit{-}}n_{\lit{+}}}$ a posteriori and the value of $\mathsf{N}$. But, for now, let us compute the steady state solution of this problem with these hypotheses

\begin{equation}
\mathsf{-\frac{1}{r}\, d_{r}(r \, \mu_{\lit{-}}\, E \, n_{\lit{-}})=N(r)=N_{0}} 
\end{equation}
\begin{equation}
\mathsf{\frac{1}{r}\, d_{r}(r \, \mu_{\lit{+}} \, E \, n_{\lit{+}})=N(r)=N_{0}}
\end{equation}
\begin{equation}
\mathsf{\frac{1}{r}\, d_{r}(r \, E)=\frac{e}{\epsilon_{\lit{0}}}(n_{\lit{+}}-n_{\lit{-}})}
\end{equation}  Since the voltage is maintained constant we have an integral condition on the electric field whose structure depends on the space charge distribution (4)

\begin{equation}
\mathsf{\int_{r_{a}}^{r_{c}}\, E(r)\, dr=V_{p}}
\end{equation}  We need two boundary conditions on $\mathsf{n_{\lit{+}}}$ and $\mathsf{n_{\lit{-}}}$. $\mathsf{Ar^{\lit{+}}}$ drift towards the cathode and $\mathsf{e^{\lit{-}}}$ towards the anode. For a steady state solution, in order to avoid an accumulation of $\mathsf{Ar^{\lit{+}}}$ at the cathode and  $\mathsf{e^{\lit{-}}}$ at the anode, $\mathsf{n_{\lit{-}}}$ and $\mathsf{n_{\lit{+}}}$ must verify
\begin{equation}
  \mathsf{n_{\lit{+}}(r_{a})=0}
\end{equation}
\begin{equation}
  \mathsf{n_{\lit{-}}(r_{c})=0}
\end{equation} The system of non linear differential equations (2) to (4) with conditions (5) to (7) is self-sufficient. They induce
\begin{equation}
  \mathsf{n_{\lit{-}}(r)=\frac{N_{0}(r_{c}^{2}-r^{2})}{2r\mu_{\lit{-}}E}}
\end{equation}
\begin{equation}
  \mathsf{n_{\lit{+}}(r)=\frac{ N_{0}(r^{2}-r_{a}^{2})}{2r\mu_{\lit{+}}E}}
\end{equation}
Thus, thanks to (4) $$\mathsf{\frac{1}{r}d_{r}((rE)^{2})= N_{0}\frac{e}{\epsilon_{0}}\biggl(\frac{(r^{2}-r_{a}^{2})}{\mu_{\lit{+}}}+\frac{(r^{2}-r_{c}^{2})}{\mu_{\lit{-}}}\biggr)}$$ We can integrate this differential equation and then we get
$$\mathsf{E(r,N_{0}, C)=[N_{0}\frac{e}{4\epsilon_{0}}\biggl(\frac{(r^{2}-2r_{a}^{2})}{\mu_{\lit{+}}}+\frac{(r^{2}-2r_{c}^{2})}{\mu_{\lit{-}}}\biggr)+\frac{C}{r^{2}} ]^{\frac{1}{2}}}$$
where $\mathsf{C}$ is a constant of integration whose value can be calculated with the condition (5). We notice that we cannot find a $\mathsf{C}$ for every $\mathsf{V_{p}}$. This is what we expected. For a given geometry of the chamber and for a given value of $\mathsf{N_{0}}$, there is a minimum voltage $\mathsf{V_{min}(N_{0})}$ under which there is not any steady state solution for the problem. The value of $\mathsf{C=C_{min}}$ for this minimum voltage is given by the condition $\mathsf{E(r_{a})=0}$. For every $\mathsf{C>C_{min}}$, the electric field is nonzero, which is a necessary condition for $\mathsf{n_{\lit{-}}}$ and $\mathsf{n_{\lit{+}}}$ to exist (see (9) and (10)). $$\mathsf{C_{min}( N_{0})=N_{0}\frac{e}{4\epsilon_{0}}r_{a}^{2}\biggl(\frac{r_{a}^{2}}{\mu_{\lit{+}}}+\frac{2r_{c}^{2}-r_{a}^{2}}{\mu_{\lit{-}}}\biggr)}$$ Thus we can say that the begining voltage of the saturation plateau is greater or equal than 
$$\mathsf{V_{min}(N_{0})=\int_{r_{a}}^{r_{c}}E(r,C_{min})\,dr}$$ whatever $\alpha$ is (the recombination constant) we cannot find a saturation below this limit. We can notice the importance of space charge effect at the begining of the saturation plateau.

\subsection{The saturation domain on a voltage-flux graph}

\subsubsection{Recombination limit : $\mathsf{V_{rec}}$}

The saturation current corresponds to the collection of all electrons created by the primary ionisation. So, $\mathsf{I_{sat}}$ is equal to $\mathsf{e\int^{r_{c}}_{r_{a}}N(r)2\pi\,r\,L\,dr}$, that is to say  
\begin{equation}
\mathsf{I_{sat}=\pi\,e\,LN_{0}(r_{c}^{2}-r_{a}^{2})}
\end{equation} We can notice that $\mathsf{I_{sat}}$ depends on the product $\mathsf{(r_{c}-r_{a})(r_{c}+r_{a})}$ whereas for a chamber for which $\mathsf{(r_{c}-r_{a})\gg r_{a}}$ , $\mathsf{N(r)}$ would be $\mathsf{\beta/r}$ and thus $\mathsf{I_{sat}}$ would depend only on $\mathsf{(r_{c}-r_{a})}$. The limitations of the saturation plateau are due to the {\it recombination} and the {\it threshold condition}. The recombination modifies the saturation current by the following quantity $$\mathsf{\delta I_{rec}=e\,\int_{V}(\alpha n_{\lit{+}}n_{\lit{-}})d\tau}$$ This quantity can be evaluated by replacing $\mathsf{n_{\lit{+}}}$ and $\mathsf{n_{\lit{-}}}$ by their value calculated in (4.1). Since the value of $\mathsf{\alpha}$ is very weak we can consider that $\mathsf{\delta I_{rec}}$ becomes important, relatively to $\mathsf{I_{sat}}$, at the vicinity of $\mathsf{V_{min}(N_{0})}$. Indeed, at this point $\mathsf{E(r_{a})=0}$, so $\mathsf{n_{\lit{+}}(r_{a})}$ and  $\mathsf{n_{\lit{-}}(r_{a})}$ tend to infinity (theoretically speaking). So is $\mathsf{\delta I_{rec}}$. That is why we assume that the beginning voltage of the saturation plateau $\mathsf{V_{rec}}$ is equal to $\mathsf{V_{min}(N_{0})}$.

\subsubsection{Threshold limit : $\mathsf{V_{max}}$}

Suppose that over a certain electric field threshold ($\mathsf{E_{S}}$) there is an electronic multiplication that ends the saturation plateau. We want to determine, for a given value of $\mathsf{N_{0}}$, the minimum voltage corresponding to the beginning of this phenomenon. It appears when either $\mathsf{E(r_{a})}$ or $\mathsf{E(r_{c})}$ becomes greater than $\mathsf{E_{S}}$ since the function $\mathsf{E(r)}$ is concave.
\\
Suppose $\mathsf{N_{0}=0}$. There is no ion-pair created in the volume of the gas. So there is no space charge problem. Thus, the electric field has the well known form in $\mathsf{1/r}$ (see graph. 1 on fig. 5). We reach the threshold condition if $\mathsf{E(r_{a})=E_{S}}$ that is to say $\mathsf{C_{1}=E_{S}^{2}r_{a}^{2}}$ and then $\mathsf{V_{max}(0)=\int E(r)dr=E_{S}r_{a}\int dr/r}$. 
\\
Suppose now that we increase the value of $\mathsf{N_{0}}$. Since ion-pairs are created in the gas, space charge may cause electric field distortion. On fig. 5 , we can see the evolution of the electric field with $\mathsf{N_{0}}$ (the voltage is maintained constant). We can meet three cases at the threshold condition :  either $\mathsf{E(r_{a})>E(r_{c})}$ and then $\mathsf{E(r_{a})=E_{S}}$ (graph. 1 on fig. 5), $\mathsf{E(r_{a})=E(r_{c})}$ and then $\mathsf{E(r_{a})=E(r_{c})=E_{S}}$ (graph. 2 on fig. 5), or $\mathsf{E(r_{a})<E(r_{c})}$ and then $\mathsf{E(r_{c})=E_{S}}$ (graph. 3 on fig. 5). The first case involves that $$\mathsf{C_{1}=E_{S}^{2}r_{a}^{2}+\frac{e}{4\epsilon_{0}}N_{0}\,r_{a}^{2}\biggl(\frac{r_{a}^{2}}{\mu_{\lit{+}}}+\frac{(2r_{c}^{2}-r_{a}^{2})}{\mu_{\lit{-}}}\biggr)}$$ and the corresponding voltage $$\mathsf{V_{max1}(N_{0})=\int^{r_{c}}_{r_{a}} E(r,N_{0},C_{1})\,dr}$$ The second and third case involve that $$\mathsf{C_{2}=E_{S}^{2}r_{c}^{2}-\frac{e}{4\epsilon_{0}}N_{0}\,r_{c}^{2}\biggl(\frac{(r_{c}^{2}-2r_{a}^{2})}{\mu_{\lit{+}}}-\frac{r_{c}^{2}}{\mu_{\lit{-}}}\biggr)}$$ and the corresponding voltage $$\mathsf{V_{max2}(N_{0})=\int^{r_{c}}_{r_{a}} E(r,N_{0},C_{2})\,dr}$$ We can notice, because of the expression of $\mathsf{C_{1}}$ and $\mathsf{C_{2}}$, that space charge  becomes important for high value of $\mathsf{N_{0}}$. Moreover this space charge effect is responsible for the maximum value of $\mathsf{N_{0 max}}$ over which saturation cannot exist.

\begin{figure}[!th]
\epsfxsize=7cm
\epsfysize=7cm
\centerline{\epsfbox{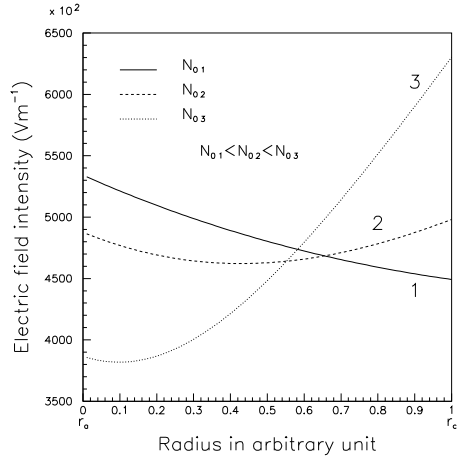}}
\caption{Electric field for various values of $\mathsf{N_{0}}$ at constant voltage.}
\label{elec}
\end{figure}

\subsection{The use of these results}

Suppose we want to know the saturation domain of a given fission chamber for a given incoming neutron spectra. We have to calculate $\mathsf{\chi=\int\sigma(\epsilon)\Phi_{\epsilon}(\epsilon)d\epsilon}$. For this purpose we can use a Monte Carlo code, like MCNP, developed by the Los Alamos National Laboratory, to simulate the collision of neutrons with the fission chamber and the fissile product inside, and then to calculate $\mathsf{\chi}$ and $\mathsf{N_{0}}$. Then, knowing the geometry of the chamber, we can get $\mathsf{V_{rec}(N_{0})}$, which is the begining voltage of the saturation plateau, and $\mathsf{V_{max}(N_{0})}$, which is the end of the saturation plateau. We can sum these results up in a graph. The abscissa is the voltage and the ordinate is either $\mathsf{N_{0}}$ or the current or the thermal flux of neutrons, depends on what we want to know. On fig.6 and fig.7 we draw various saturation domains, each being limitated by two curves. These curves represent $\mathsf{V_{rec}}$ and $\mathsf{V_{max}}$. We wrote a very simple code that permits us to generate the graph that represents $\mathsf{V_{min}(N_{0})}$ and $\mathsf{V_{max}(N_{0})}$. The inputs of this code are : $\mathsf{r_{a}}$, $\mathsf{r_{c}}$, $\mathsf{L}$ (geometry) and $\mathsf{\mu_{\sigma}}$, $\mathsf{M_{f}}$ (fissile product). The outputs are three arrays of points : $\mathsf{V_{min}(N_{0})}$, $\mathsf{V_{max}(N_{0})}$ and $\mathsf{N_{0}}$.

\section {MODEL VERSUS TECHNICAL DATAS}

We want to compare the results we get by the use of our model with the characteristic of two industrial fission chambers manufactured by \textsf{PHOTONIS, imaging sensors} : \textsf{CFUR 43} and \textsf{CFUF 43}. Let us begin by the first one. The table 1 gives us the geometrical and physical features of the chamber.
\\
We put $\mathsf{r_{c}}$, $\mathsf{L}$, pressure  and $\mathsf{\mu_{\sigma}}$ to their actual value in our code. \textsf{PHOTONIS, imaging sensors} writes in the \textsf{CFUR43} manual that the optimum polarisation is \textsf{150 V}. Let us say that we do not know $\mathsf{r_{a}}$. We want to find its value with the help of our code. We draw on the following graph (fig. 6) various domain of saturation of \textsf{CFUR 43} for various value of $\mathsf{r_{c}-r_{a}}$ (inter-electrode gap). We can remark that values of $\mathsf{r_{c}-r_{a}}$ between \textsf{0.2 mm} and \textsf{0.3 mm} fit with the requirement of a \textsf{150 V} optimum polarisation. That is exactly\footnote{the exact value of $\mathsf{r_{a}}$ is an industrial secret of \textsf{PHOTONIS}} the range value given by \textsf{PHOTONIS, imaging sensors}

\begin{figure}[!th]
\epsfxsize=7cm
\epsfysize=7cm
\centerline{\epsfbox{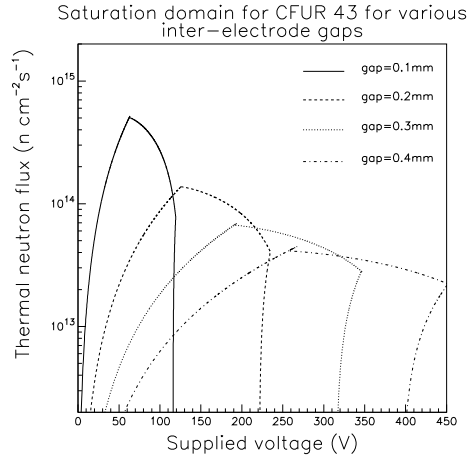}}
\caption{Variation of the saturation domain with the inter-electrode gap for \textsf{CFUR} chamber}
\label{cfur}
\end{figure}

You could say that it only works for \textsf{CFUR 43}. So let us see it works with another chamber with different size : \textsf{CFUF 43} (see table 2). The result given by our model can be visualised on fig. 7.

\begin{figure}[!th]
\epsfxsize=7cm
\epsfysize=7cm
\centerline{\epsfbox{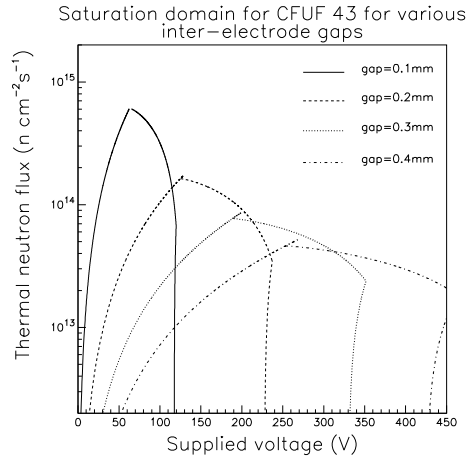}}
\caption{Variation of the saturation domain with the inter-electrode gap for \textsf{CFUF} chamber.}
\label{cfuf}
\end{figure}

We can also compare the sensitivity of these fission chambers (given by \textsf{PHOTONIS}) with the one calculated with our model. The sensitivity is the ratio of the current delivered by the chamber and the thermal neutron flux $$\mathsf{S=\frac{I_{sat}}{\Phi_{th}}=\frac {\pi^{2}}{2}eLI_{0}^{c}p\frac{\mu_{\sigma}}{M_{U}}N_{A}\sigma_{th}(r_{c}^{2}-r_{a}^{2})}$$ The various results are summarized in table 3. Tabulated and calculated values are in very good agreement. We can deduce from these two figures that an increase of $\mathsf{(r_{c}-r_{a})}$ induces a decrease of the maximum flux detectable. It also induces an increase of the value of the optimum voltage for saturation, and of the width of the saturation plateau.

\section{MODEL VERSUS SILOE RESULTS}

In this part we present the $\mathsf{1.5~mm}$ diameter subminiature fission chamber manufactured at the \textsf{CEA-Cadarache, France} (see table 4). We tested two home made $\mathsf{1.5~mm}$ diameter chambers : a $\mathsf{1.1~bar}$ and a $\mathsf{4.0~bar}$ chamber (this is the pressure of argon in the chamber). These experiments were carried out at the \textsf{CEA/Gre\-no\-ble (France)} at the \textsf{SILOE} facility (experimental nuclear power plant). The aim of these experiments was, first, to make some general empirical observations, and second, to collect enough data to validate our model.

\subsection{Comparison between a $\mathsf{1.1~bar}$ and a $\mathsf{4.0~bar}$ fission chamber}

We compared the characteristics of both fission chambers in the same thermal flux of about $\mathsf{3.4\,10^{12}n.cm^{-2}.s^{-1}}$. We can see in fig. 8, as expected, that the saturation current for a $\mathsf{4.0~bar}$ fission chamber is about four times greater than the one for a $\mathsf{1.1~bar}$ chamber. Actually, the accurate ratio is $\mathsf{3.5}$. But we can also notice that the saturation plateau, for a $\mathsf{1.1~bar}$ chamber, begins at a voltage lower than the one for a $\mathsf{4.0~bar}$ chamber. 
\begin{figure}[!th]
\epsfxsize=7cm
\epsfysize=7cm
\centerline{\epsfbox{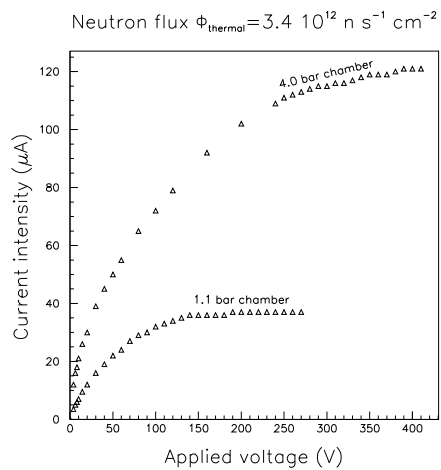}}
\caption{Characteristic of two chambers for the same neutron flux.}
\label{twofc}
\end{figure}
Thus, we can deduce from those remarks the following rule : {\bf if, for a given neutron flux, the voltage to reach the saturation plateau is too high, we can reduce the argon pressure in the chamber, but as a consequence, we also reduce the saturation current}. This rule is verified with our model. On fig. 9 and fig. 10 one can see the superposition of experimental points and of $\mathsf{V_{rec}}$ (line representing the beginning of the saturation domain). On these figures, ``?'' means that we didn't know the thermal neutron flux at this location in the core. Once more, the model is in good agreement with the experimental data. It predicts with a good accuracy the beginning of the saturation domain.
\begin{figure}[!ht]
\epsfxsize=7cm
\epsfysize=7cm
\centerline{\epsfbox{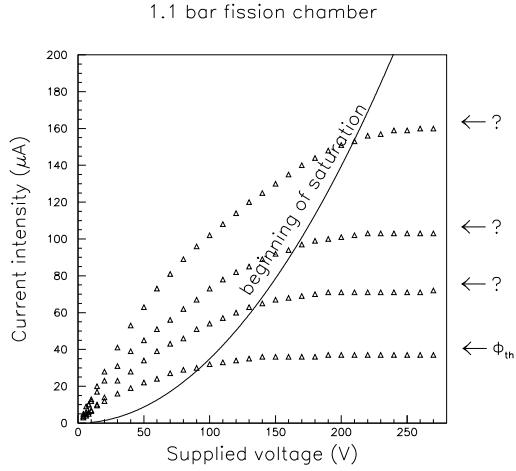}}
\caption{Characteristic of a $\mathsf{1.0~bar}$ chamber for various neutron fluxes.}
\label{onebar}
\end{figure}

\begin{figure}[!ht]
\epsfxsize=7cm
\epsfysize=7cm
\centerline{\epsfbox{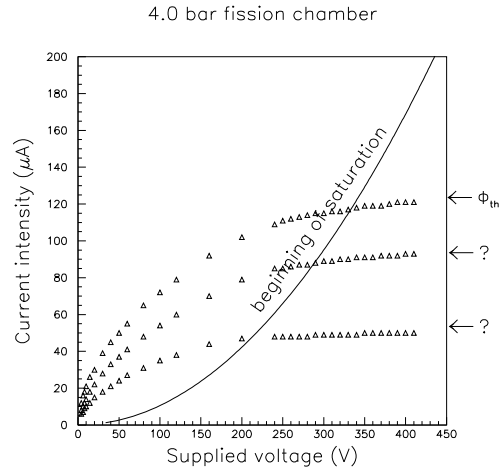}}
\caption{Characteristic of a $\mathsf{4.0~bar}$ chamber for various neutron fluxes.}
\label{fourbar}
\end{figure}
For a $\mathsf{1.1~bar}$ chamber we measured characteristics up to $\mathsf{270~V}$. Above this voltage the current intensity increased very quickly. That is why we did not measure further in order not to destroy the chamber. This is the same reason why we did not carry out the experiment above $\mathsf{430~V}$ for the $\mathsf{4.0~bar}$ chamber. We calculated the sensitivity of each chamber with measured data (for $\mathsf{\Phi_{th}=3.4\,10^{12}n.cm^{-2}.s^{-1}}$, the $\mathsf{1.1bar}$ chamber gives $\mathsf{I_{sat}=37\mu A}$ and the $\mathsf{4.0bar}$ chamber gives $\mathsf{I_{sat}=120\mu A}$) and with our model. The results are summarized in table 5. The order of magnitude is the same. The discrepancy between measured and calculated values is consistent with the uncertainty of the pressure and geometry.

\section{CONCLUSION}

The model described in this paper is in good agreement with technical data for existing fission chambers \textsf{CFUR 43} and \textsf{CFUF 43}, and also with experimental data obtained with home made subminiature fission chambers. The model also takes into account the temperature, which is necessary for the use of a fission chamber in the core of a nuclear power plant. Finally, this study revealed two important behaviors of fission chamber : firstly, in order to reduce the value of the minimum voltage of saturation, we can reduce the pressure in the chamber, but, as a consequence, we reduce the sensitivity of the chamber. Secondly, we learned that increasing the inter-electrode gap leads to a decrease of the maximum flux detectable and an increase of the optimum voltage of saturation and of the width of the saturation plateau.

\listoffigures

\eject

\begin{table}
\caption{Geometrical and physical features of \textsf{CFUR 43}.}
\vspace{.5 cm}
\begin{tabular}{|lr|}
\hline
\multicolumn{2}{|c|}{\textsf{CFUR 43 Features}}
\\ \hline\hline
$\mathsf{r_{c}}$ & $\mathsf{1.25~mm}$
\\ \hline
$\mathsf{L}$ & $\mathsf{13~mm}$ 
\\ \hline
$\mathsf{pressure}$ & $\mathsf{1.1~bar}$ 
\\ \hline
$\mathsf{\mu_{s}}$ & $\mathsf{200~\mu g\cdot cm^{-2}}$
\\ \hline
\end{tabular}
\end{table}

\begin{table}
\caption{Geometrical and physical features of \textsf{CFUF 43}.}
\vspace{.5 cm}
\begin{tabular}{|lr|}
\hline
\multicolumn{2}{|c|}{\textsf{CFUF 43 Features}}
\\ \hline\hline
$\mathsf{r_{c}}$ & $\mathsf{1.78~mm}$
\\ \hline
$\mathsf{L}$ & $\mathsf{27~mm}$ 
\\ \hline
$\mathsf{pressure}$ & $\mathsf{1.1~bar}$ 
\\ \hline
$\mathsf{\mu_{s}}$ & $\mathsf{160~\mu g\cdot cm^{-2}}$
\\ \hline
\end{tabular}
\end{table}

\vspace{.5 cm}
\begin{table}
\caption{Comparaison of tabulated and calculated sensitivity for \textsf{CFUR 43} and \textsf{CFUF 43}.}
\vspace{.5 cm}
\begin{tabular}{|l|r|r|}
\hline
\multicolumn{3}{|c|}{\textsf{Sensitivity of fission chambers}}
\\ \hline\hline
      & $\mathsf{CFUR 43}$ & $\mathsf{CFUF 43}$
\\ \hline
$\mathsf{S_{tabulated}(A/n.cm^{-2}.s^{-1})}$ & $\mathsf{3.\,10^{-18}}$ & $\mathsf{1.\,10^{-17}}$
\\ \hline
$\mathsf{S_{calculated}(A/n.cm^{-2}.s^{-1})}$ & $\mathsf{3.3\,10^{-18}}$ & $\mathsf{0.9\,10^{-17}}$
\\ \hline
\end{tabular}
\end{table}

\begin{table}
\caption{Geometrical and physical features of subminiature fission chamber.}
\vspace{.5 cm}
\begin{tabular}{|lr|}
\hline
\multicolumn{2}{|c|}{\textsf{Subminiature Fission Chamber Features}}
\\ \hline\hline
$\mathsf{anode~radius~~r_{a}}$ & $\mathsf{0.35~mm}$
\\ \hline
$\mathsf{cathode~radius~~r_{c}}$ & $\mathsf{0.65~mm}$
\\ \hline
$\mathsf{length~of~sensitive~part~~L}$ & $\mathsf{12~mm}$ 
\\ \hline
$\mathsf{pressure~of~argon}$ & $\mathsf{1.1~bar + 4.0~bar}$ 
\\ \hline
$\mathsf{surfacic~mass~of~fissile~product~~\mu_{s}}$ & $\mathsf{760~\mu g\cdot cm^{-2}}$
\\ \hline
\end{tabular}
\end{table}

\vspace{.5 cm}
\begin{table}
\caption{Comparaison of measured and calculated sensitivities for 1.1 bar and 4.0 bar subminiature fission chambers.}
\vspace{.5 cm}
\begin{tabular}{|l|r|r|}
\hline
\multicolumn{3}{|c|}{\textsf{Sensitivity of subminiature fission chambers}}
\\ \hline\hline
      & $\mathsf{1.1~bar~chamber}$ & $\mathsf{4.0~bar~chamber}$
\\ \hline
$\mathsf{S_{measured}(A/n.cm^{-2}.s^{-1})}$ & $\mathsf{1.1\,10^{-17}}$ & $\mathsf{3.6\,10^{-17}}$
\\ \hline
$\mathsf{S_{calculated}(A/n.cm^{-2}.s^{-1})}$ & $\mathsf{0.7\,10^{-17}}$ & $\mathsf{2.8\,10^{-17}}$
\\ \hline
\end{tabular}
\end{table}

\end{document}